\documentclass[twocolumn]{aastex631}

\newcommand{\Ssfr}        {\mbox{$\Sigma_{\rm SFR}$}}
\newcommand{\Sstar}       {\mbox{$\Sigma_{\star}$}}
\newcommand{\Smol}       {\mbox{$\Sigma_{\rm mol}$}}

\newcommand{\EW}         {\mbox{${\rm EW(H\alpha)}$}}
\newcommand{\Mstar}      {\mbox{$M_\mathrm{{\star} }$}}
\newcommand{\Mmol}      {\mbox{$M_\mathrm{{mol} }$}}
\newcommand{\fg}      {\mbox{$\mathrm{f_{mol} }$}}

\newcommand{\Dsfms}      {\mbox{$\mathrm{\Delta SFMS}$}}

\newcommand{\Dsk}      {\mbox{$\mathrm{\Delta SK}$}}

\newcommand{\Dmgms}      {\mbox{$\mathrm{\Delta MGMS}$}}
\newcommand{\Ha}          {\mbox{$\mathrm{H\alpha}$}}

\newcommand{\Reff}          {\mbox{$\mathrm{R_{eff}}$}}
\newcommand{\Msun}          {\mbox{$\mathrm{M_{\odot}}$}}

\newcommand{\CO}          {\mbox{$\mathrm{^{12}CO}$}}
\newcommand{\COIII}          {\mbox{$\mathrm{^{13}CO}$}}

\newcommand{\pcmc}          {\mbox{$\mathrm{pc^{-2}}$}}

\newcommand{\Wt}          {\mbox{$\mathrm{W_T}$}}

\graphicspath{{./}{figures/}}
\DeclareUnicodeCharacter{0301}{\'{e}} 

\begin{document}

\title{Exploring the Impact of Galactic Interactions and Mergers on the Central Star Formation of APEX/EDGE-CALIFA Galaxies}

\author[0000-0002-1135-7043]{Yeny Garay-Solis}
\author[0000-0003-2405-7258]{Jorge K. Barrera-Ballesteros}
\affiliation{Instituto de Astronom\'ia, Universidad Nacional Aut\'onoma de M\'exico, A.P. 70-264, 04510 M\'exico, D.F., M\'exico}

\author[0000-0001-6498-2945]{Dario Colombo}
\affiliation{Argelander-Institut für Astronomie, Auf dem Hügel 71, 53121 Bonn, Germany}

\author[0000-0001-6444-9307]{Sebastián F. Sánchez}
\affiliation{Instituto de Astronom\'ia, Universidad Nacional Aut\'onoma de M\'exico, A.P. 70-264, 04510 M\'exico, D.F., M\'exico}

\author[0000-0001-9226-9178]{Alejandra Z. Lugo-Aranda}
\affiliation{Instituto de Astronom\'ia, Universidad Nacional Aut\'onoma de M\'exico, A.P. 70-264, 04510 M\'exico, D.F., M\'exico}

\author[0000-0002-5877-379X]{Vicente Villanueva}
\affiliation{Department of Astronomy, University of Maryland, College Park, MD 20742, USA}

\author[0000-0002-7759-0585]{Tony Wong}
\affiliation{Department of Astronomy, University of Illinois, Urbana, IL 61801, USA}

\author[0000-0002-5480-5686]{Alberto D. Bolatto}
\affiliation{Department of Astronomy, University of Maryland, College Park, MD 20742, USA}

%\affiliation{Instituto de Astronomı́a, Universidad Nacional Autónoma de México, A.P. 70-264, 04510 México, D.F., Méxic}

%\collaboration{6}{(AAS Journals Data Editors)}

%\author{Butler Burton}
%\affiliation{Leiden University}
%\affiliation{AAS Journals Associate Editor-in-Chief}

%\altaffiliation{AASTeX v6+ programmer}

%\affiliation{AAS Director of Publishing}
%\affiliation{American Astronomical Society \\
%1667 K Street NW, Suite 800 \\

%\author{Magaret Donnelly}
%\affiliation{IOP Publishing, Washington, DC 20005}



\begin{abstract}

Galactic interactions and subsequent mergers are a paramount channel for galaxy evolution. %In this paper, we study how the specific star formation rate, the molecular gas fraction and the star formation efficiency behave in the central region of galaxies (approximately within an effective radius) at different merger stages.
In this work, we use the data from 236 star forming CALIFA galaxies with integrated molecular gas observations in their central region (approximately within an effective radius) -- from the APEX millimeter telescope and the CARMA millimeter telescope array. This sample includes isolated (126 galaxies) and interacting galaxies in different merging stages (110 galaxies; from pairs, merging and post-merger galaxies). 
%This is the first work that studies the impact of mergers and interactions in a sample with homogeneous observations, both in the visible and radio spectrum, and significantly large. 
We show that the impact of interactions and mergers in the center of galaxies is revealed as an increase in the fraction of molecular gas (compared to isolated galaxies). %However, the specific star formation rate and the star formation efficiency show a decrease. %
Furthermore, our results suggest that the change in star formation efficiency is the main driver for both an enhancement and/or suppression of the central star formation -- except in merging galaxies where the enhanced star formation appears to be driven by an increase of molecular gas. We suggest that gravitational torques due to the interaction and subsequent merger transport cold molecular gas inwards, increasing the gas fraction without necessarily increasing star formation.

\end{abstract}


\section{Introduction} 
\label{sec:intro}

The evolutionary processes that a galaxy goes through affect its internal properties. Among the most significant processes are: star formation, gas recycling, metallicity enrichment, and the energy produced by supernovae or in active nuclei \citep{Kormendy2004,Kormendy2013}. An external and fast process that significantly affects the evolution of galaxies on a large scale is galactic interactions/mergers \citep{Toomre1977,Conselice2003,Christensen2009,Kormendy2013}. A large number of observational and numerical studies have suggested an enhancement of star formation in galactic centres, related to the increase in molecular gas in these objects due to the tidal forces exerted by the galaxies that undergo this merger \citep[e.g., ][]{Patton1997,Patton2005,Smith2007,Li2008,Ellison2008,Ellison2010,Patton2011,Ellison2011,Scudder2012,Patton2013,Ellison2013,Pan2018,Moreno2021,Ueda2021}. However, recent studies using sub-millimeter  observations have suggested that rather than an enhancement in the central star formation rate (SFR), the interaction tend to increase the molecular gas fraction, i.e. the molecular gas mass-over-stellar mass ratio \citep[e.g., ][]{Kaneko2022}. Given the lack of a statistically robust sample with homogeneous observations at both optical and sub-millimeter wavelengths, it has been difficult to reliably quantify the impact of the merging process in shaping the central star formation in galaxies. 

%On the other hand, \citet{Kaneko2022} study of four pairs of interacting/merging galaxies with kpc-resolved data, and molecular gas measurements, found that most of their galactic-scale sample did not exhibit a significant increase in their rate of star formation, compared to the best fit of isolated galaxies. Their studies suggest that the molecular gas does not vary in the early stages of the interaction, although their interacting galaxies have a higher molecular gas fraction than their isolated galaxies. However, those studies that seek to estimate the variation in the amount of molecular gas have small samples, which makes it difficult to quantify the real impact of interactions and mergers in the gas fraction, as well as star formation in the center of the galaxies.

In this study, we explore how the specific star formation rate (sSFR), molecular gas fraction (\fg), and star formation efficiency (SFE) behave in the centre of 236 Calar Alto Legacy Integral Field Area (CALIFA) survey \citep{califa2012} star-forming galaxies at different merging stages compared to isolated galaxies. Our sample includes 140 galaxies  from the Atacama Pathfinder EXperiment (APEX) submillimeter telescope \citep{apex2006}, first presented by \citet{Colombo2020}, and 96 galaxies with observations obtained by the Combined Array for Research in Millimeter-wave Astronomy (CARMA) \citep{carma2006}, encompassed by the  Extragalactic Database for Galaxy Evolution (EDGE-CALIFA) survey \citep{Bolatto2017}. These observations provide integrated measurements from the central region covering approximately one effective radius ($\sim$1~\Reff). This is the largest CO database with optical Integral Field Spectroscopy (IFS) data so far.
%This article is organized as follows:
This article has the following structure: 
in Sec.~\ref{sec:data_analysis} we describe the data and sample, %Sect.~\ref{subsec:data_califa} we present the optically derived parameters of CALIFA survey and in Sect.~\ref{subsec:data_apex_carma} CO-derivatives of APEX and CARMA,
%in Sect.~\ref{subsec:derived_parameters} we describe how the parameters used in this study are derived,
while in Sec.~\ref{sec:Results} we present our results of the analyzes carried out at each merger stage sample. In Sec.~\ref{sec:discusion} we present discussion of our results. Finally, we summarize our study and present the conclusions of the results in Sec.~\ref{sec:conclusions}.
%Sect.~\ref{subsec:morphological_classification} we define the morphological classification of our merger stages, we describe the analysis made to the sample to characterize the physical properties of the SFG in Sect.~\ref{subsec:characterizing_SFG}, in Sect.~\ref{subsec:mock_data} we explain how we perform scaling relations by generating random data. In 
For the derived data, we use a cosmology with the parameters: $\mathrm{H_0=71\,km\,s^{-1}\,Mpc^{-1}}$, $\mathrm{\Omega_M=0.27}$, $\mathrm{\Omega_\Lambda=0.73}$.

\section{Data, Sample  and Analysis} 
\label{sec:data_analysis}

\subsection{CALIFA data}
\label{subsec:data_califa}

In this study, we use galaxies included in the {Calar Alto Legacy Integral Field Area} (CALIFA) sample \citep{califa2012}. This survey observed more than 600 galaxies in the local Universe ($0.005<z<0.03$) using Integral-Field Spectroscopy (IFS). CALIFA used the Integral Field Unit (IFU) PPAK \citep{PPAK2006}, which was attached to the 3.5~m telescope, with 74~$\times$~64 arcsec of Field of View (FoV) of the Calar Alto Observatory in Spain. For each observed galaxy, this FoV covers $\sim2.5~\Reff$ with a typical spatial physical resolution of $\sim$1~kpc. 
%The PPAK instrument was built using the fiber optic bundle technique. Each fiber has a spatial aperture projected skywidth of $\sim$2.7 arcsec, thus, for the redshift of the sample ($0.005 < z < 0.03$), it provides a spatial physical resolution of $\sim$1~kpc. 
The spectrograph used was the PMAS ({Potsdam Multi Aperture Spectrograph}) \citep{PMAS2005}. This spectrograph has two superimposed grid configurations: V500 (3750-7000~\AA) with a (low) spectral resolution of 6~\AA; and V1200 (3700-4700~\AA) with an (intermediate) spectral resolution of 2.7~\AA.
%which is used for stellar absorption and kinematic analyzes of gas and stars \citep{califa2012}. 
The raw data cube obtained from PMAS/PPAK was reduced with a specialized pipeline for the CALIFA survey \citep{Califa_reduccion}. The reduced data cubes were analyzed with PIPE3D \citep{pipe3D1,pipe3D2} to derive physical parameters of the stellar populations and emission lines of each spaxel for a galaxy. The reduced data cubes of 667 galaxies are publicly available (CALIFA DR3 \footnote{https://califa.caha.es/}).  
%The most relevant parameters for the current study the parameters used in this study
%The parameters used  are the stellar mass,  the rate of star formation and the equivalent width of \Ha\, (\EW), which were derived from the PIPE3D analysis. For more details of the analysis process see \citet{Review2021}.
Below we present the integrated derived parameters from these data cubes.

\subsection{Molecular gas mass }
\label{subsec:data_apex_carma}

%The molecular gas mass (\Mmol) was estimated by the relation between the integrated intensity of the observed CO line, the molecular gas density and a conversion factor \citep{BolattoReview}. 
We used one of the largest databases of CO observations available for galaxies with IFS data in order to derive the central molecular gas mass (\Mmol). 
Our sample includes measurements made by the {Atacama Pathfinder EXperiment} (APEX) submillimetre telescope  \citep{apex2006} and {Combined Array for Research in Millimeter-wave Astronomy} (CARMA)  \citep{carma2006}. 
This data and its calibration are described in detail in \citet{Colombo2020} and \citet{Bolatto2017}, respectively; here, we summarize their main features.
%
%which are detailed in \citet{Colombo2020}, where observation, data reduction, calibration and procedure to transform CO intensities into mass of molecular gas are explained.
%
%APEX has a 12 meter diameter primary reflector and operates at millimeter and submillimeter wavelengths, that is, between far infrared light and radio waves \citep{apex2006}. The measurement data is calibrated and reduced using the GILDAS software (\textit{Grenoble Image and Line Data AnalysiS}), generating the measurement of \Lco\, ($\mathrm{K\,km\,s^{- 1}}$) \citep{apex2006,apexLASMA}. 
%

\citet{Colombo2020} presented APEX measurements of the \CO(2-1) line ($\nu=230.538$~GHz) for a sample of 296 galaxies included in the CALIFA survey. These objects were selected to 
%restricted to be located in declination $\leq 30^o$ and
have an inclination angle below $65^o$. Furthermore, all galaxies have central measurements and 39 of these objects \textbf{also} have off center measurements. The APEX beam coverage is 26.3 arcsec, so the APEX beam coverage is approximate 1.12~\Reff\ for this sample of galaxies.

On the other hand, the Extragalactic Database for Galaxy Evolution (EDGE) survey \citep{Bolatto2017} is observed a sample of galaxies drawn from the CALIFA survey using the CARMA interferometer.
EDGE provides CO maps with a sensitivity of $\Smol\sim11\,\Msun\,\pcmc $ (before tilt correction) and a physical resolution of $\sim1.4$~kpc. 
The EDGE-CALIFA survey was aimed to complement the optical information derived from the CALIFA survey with CO maps observed by CARMA. The CO observations at the transitions of \CO(1-0) at $\nu\sim$115.3 GHz and \COIII(1-0) at $\nu\sim$ 110.2 GHz, in the same coverage as the FoV of CALIFA (74 x 64 arcsec) \citep{Bolatto2017}. 

\subsection{Derived parameters}
\label{subsec:derived_parameters}
For this study, we use the SFR, the \Mstar, and the \Mmol\, estimated within the APEX beam aperture, both for APEX and CARMA galaxies. The parameters are convoluted by a \Wt\, function before being integrated \citep{Colombo2020}. The optical parameters are integrated from their angular-resolved properties (i.e., the SFR and stellar mass surface density). %. The reason behind that is that CARMA and CALIFA data are resolved and have different resolutions compared to APEX unresolved data. \citet{Colombo2020}  implement  \Wt\, to simulate the effect that the APEX beam would have on the CALIFA and CARMA data, so the parameters derived by the three data sets are comparable. 
%Briefly, the function \Wt\, is a bi-dimensional Gaussian of unit amplitude and centered at the center of the galaxy. %
%
%With the Gaussian function, a matrix of the same number of pixels of the maps (both APEX and CARMA) is created, where each element has a value according to function and the largest value is in the center. Gaussian filter of \textit{i} pixels of the same number of pixels of the maps is obtained, that is, the value of the function \Wt\ is obtained for each pixel (\Wti). 
%The stellar mass in the beam coverage  is obtained by integrating the \Sstar\, of each of the pixels of the maps ($\mathrm{\Sigma_{\star,i}}$) multiplied by the function \Wti:
%$$\mathrm{\Mstar=\sum_i\,\Sigma_{\star,i}\,\times\,\Wti}$$
%similarly for the SFR derived in the beam coverage:
%$$\mathrm{SFR=\sum_i\,SFR_{i}\,\times\,\Wti}$$
%The conversion factor (\aCO, see Eq.~\ref{eq:Mmol}) depends on the metallicity and \Mstar\, of each galaxy \citep{BolattoReview}, so the CALIFA maps were multiplied by the function \Wt. Using these maps, an \aCO\, map was generated. 
%The \aCO\, of the beam coverage was the median of the \aCO\, map.
%Finally,  the \Mmol\, of each galaxy was derived using the Eq.~\ref{eq:Mmol}, that is, the \aCO\, median was multiplied by the \Lco\, both derived by APEX and CARMA.

We derive the intensive properties by dividing each extensive property (SFR, \Mstar, \Mmol) by the area of the beam, thus correcting for the inclination angle of each galaxy. This correction is based on the ellipticity and position angle derivations of the CALIFA galaxies made by \citet{Coba2019,Lacerda2020,Sanchez2021}. Using this correction, we obtained the \Ssfr, \Sstar, and \Smol \, for each galaxy in our sample. The intensive properties of both APEX-CALIFA and EDGE-CALIFA are presented in \citet{Sanchez2021}.

Since we are interested only in star-forming galaxies, we select those galaxies with an H$\alpha$ equivalent width  ($\EW$) larger than 6~\AA\, (measured within the beam) \citep[see][for more details]{SanchezAnnRev2020}. Furthermore, using the emission line ratio diagnostics (also known as BPT diagram) \citep{Baldwin1981}, we  select galaxies that fall below the Kewley demarcation line \citep{Kewley2001}. Finally, we also consider galaxies with central reliable measurements of the CO flux, these are measurements with a signal-to-noise ratio greater than 3. These selection criteria yield a final  sample of 236 galaxies. 
%We discard the APEX and EDGE galaxies that present a $\EW~<~6$~\AA, since it is more characteristic of galaxies with little star formation \citep{SanchezAnnRev2020}.
It consists mostly of Sa to Sd galaxies and few Irr, E, and S0 (230 and 6 galaxies respectively). Also, this sample covers a wide range of stellar masses ($\mathrm{10^{8.3}-10^{11.0}\,\Msun }$), SFRs ($\mathrm{10^{-1.8}-10^{0.4}\,\Msun\,yr^{-1}}$) and molecular mass ($\mathrm{10^{5.8}-10^{10.0}}\,\Msun$) within the APEX beam coverage.

\subsection{Classification of the merger stages}
\label{subsec:morphological_classification}

%We use a morphological classification scheme for galactic merger stages. This scheme was first introduced by \citet{Veilleux2002}  to analyze a sample of local ultraluminous infrared galaxies (ULIGs).
To gauge the interaction/merging stage, we use the  classification scheme introduced by \citet{Veilleux2002}. This scheme was based on numerical simulations of a spiral-spiral merger, tracing different snapshots of that merger \citep{Surace1998}. This classification has been successfully used to characterize the merging stage for a sub-sample of CALIFA galaxies \citep{Barrera2015}. Following this study, here we classify galaxies according to their features observed in the $r$-band Sloan Digital Sky Survey  \citep[SDSS,][]{York_2000} at each merging stage. 
%
%This classification is seen as a set of stages that a spiral galaxy goes through in its evolution process through the merger with another spiral galaxy. 
%This scheme is based on the work of \citet{Surace1998} and the results of the numerical simulations of \citep{Barnes1992,Mihos1996}. 
%Also, despite the fact that this classification is derived from numerical simulations of two spiral galaxies, \citet{Barrera2015} uses it for mergers of different morphological types. For example, the spiral galaxy fusion with an elliptical one, so it is used as a broad indicator of merging stages.
Below we briefly describe the main morphological feature to classify a galaxy in a given interacting/merging stage:

\noindent \textbf{Pre-merger stage (pre):} The galaxy has a companion, each of the galaxies presents an unperturbed morphology  (i.e., they can be associated to either a spiral or an elliptical). We further use the following criteria:  
\begin{itemize}
     \item  The projected separation between each companion is  $\mathrm{r_p<160\, kpc}$ \citep[][]{Barrera2015}. The above is  based on the results of \citet{Patton2013}, who found (in observations and  simulations) that in galaxy pair with a mass ratio of 2.5:1, an increase in star formation up to projected separations of $\sim$150~kpc.
     \item systemic velocity difference $ \Delta \mathrm{v <300\, km\, s^{-1}}$ \citep[][]{Barrera2015}, which is based on the constraints   defined for large studies of galaxy pairs in from \citet{Ellison2008,Ellison2013}.
     \item  difference in magnitudes in the $r$-band $ \mathrm{ <2\, mag}$ \citep[][]{Barrera2015}.
\end{itemize}

\noindent \textbf{Merging stage (mer):} merging galaxies with features such as tidal tails, bridges, plumes but with a  nucleus well defined in each companion.

\noindent \textbf{Post-merger stage (post):} galaxies with their nuclei merged and prominent tidal tails. The emission in the core can be spread out and narrowed by dust lanes.

\noindent \textbf{Merger remnant stage (rem):} remnant  with features such as weak tidal tails, shell shapes, and ripples but fused cores.

For our interacting sample we classify 70, 17, 13, and 10 galaxies in the pre-merger, merging, post-merger, and merger remnant stage, respectively. %\textbf{An example of each of these stages is presented in Fig.~\ref{fig:ej_stages}.}

%\begin{figure*}[ht!] 
%\centering
%    \includegraphics[width=1\linewidth]{Figura_1.png}
%    \caption{SDSS $gri$-band images of galaxies included in our sample. Each box shows an example of the merger stages described in the Sect.~\ref{subsec:morphological_classification}. We show UGC~5695 as an example of an isolated galaxy. For pre-mergers, we include IC~0208 with NGC~082. In merging galaxies, we show NGC~7253B and NGC~7253A, while for post-mergers, we include NGC~7787. Finally, as an example of a merger remnant, we show NGC~5519.  In isolated and pre-merger galaxies, the green circle represents a diameter of 10~arcmin. In the other stages, the circle has a diameter of 2~arcmin.} 
%    \label{fig:ej_stages}
%\end{figure*}

\subsection{Characterizing the physical properties of Star-Forming Galaxies}
\label{subsec:characterizing_SFG}
%\Com{Start translation by Jorge and... Yeny}

One of the main goals of this study is to shed light on how the properties related to the star-formation in the center of galaxies are affected by the merger sequence. To achieve it, we explore in this section three of the most important scaling relations related to the variation of star-formation in galaxies: (1) the  Star Formation Main Sequence (SFMS) \citep[e.g.,][]{Cano2016,Cano2019,Lin2019,SanchezAnnRev2020,Ellison2021,Sanchez2021}, (2) the Schmidt-Kennicutt law (SK) \citep[e.g.,][]{Kennicutt2012,SanchezAnnRev2020,Ellison2020,Lin2020,Morokuma2020,Ellison2021,Sanchez2021,Kaneko2022}, and (3) the Molecular Gas Main Sequence (MGMS) \citep[e.g.,][]{Lin2019,SanchezAnnRev2020,Barrera2020,Ellison2021,Sanchez2021}. 

Following \citet{Lin2019,Ellison2020,Barrera2020,Sanchez2021,Ellison2021}, we use intensive central properties, i.e., normalized to the physical area of the observed aperture (see Sect.~\ref{subsec:derived_parameters} for more details) to derive these scaling relations for our sample. To characterize these scaling relations we follow a similar procedure described in \citet{Sanchez2021}. In a nutshell, for each of the scaling relations we perform a linear fit (in logarithm scales) to the data.
These scaling relations, as well as their fits, are presented in Appendix~\ref{Appendix_A}.
The best fit parameters of these three scaling relations are in excellent agreement with those derived by  \citet{Sanchez2021}.

Using those best fits, we define the residuals for each of the explored scaling relations:
\mbox{$\Dsfms=\log(\Ssfr)-(\alpha\,\log(\Sstar)+\beta)$}, \mbox{$\Dsk=\log(\Ssfr)-(\alpha\,\log(\Smol)+\beta)$}, and \mbox{$\Dmgms=\log(\Smol)-(\alpha\,\log(\Sstar)+\beta)$} for the SFMS, SK, and MGMS relations, respectively\footnote{The $\alpha$, and $\beta$ best-fitted parameters are different depending on the explored scaling relation, see Appendix~\ref{Appendix_A}.}. 
To account for individual errors, we performed a Monte Carlo (MC) iteration by perturbing the original data within the error range and repeating the complete analysis 1000 times. The reported results in this study are the average of all MC iterations.
For each of these scaling relations, the ratio of the two parameters corresponds to another physical property (i.e., the specific SFR, \mbox{sSFR = \Ssfr/\Sstar}; the star-formation efficiency, \mbox{SFE = \Ssfr/\Smol}; and the gas fraction, \mbox{\fg = \Smol/\Sstar}). We find a significant good correlation between the ratio derived from the parameters and the residuals from the best fit ($r > 0.9$ for the three scaling relations).
%We note that this is not the case when using extensive scaling relations, since their slopes are much less than 1 (see Appendix~\ref{Appendix_A}).
%, that is, the residuals and ratios derived from the scaling relations using extensive central properties. 
This suggests that intensive properties are less prone to be affected by distance effects as they are normalized to the physical area. Our results indicate that exploring the residuals of the scaling relations using intensive properties is equivalent to explore their corresponding ratios. In Sect.~\ref{sec:Results} we use these residuals to quantify the impact of the merger sequence in the central star-forming properties. In the next section we use the best fit parameters to replicate the observed scaling relations using a sample of randomly distributed data. 

\subsection{Scaling relations using random data}
\label{subsec:mock_data}

As we mention in the previous section, we use the residuals of the aforementioned scaling relations to explore the impact of the merger sequence on the star formation stage in the central region of galaxies. Recently, the exploration of these residuals has been used to estimate the drivers of star-formation in galaxies \citep[e.g.,][]{Ellison2020,Lin2020}. On the other hand, using large samples of galaxies, different studies have also cautioned on the use of these relations between residuals or ratios as they may not have a physical but rather statistical origin \citep[e.g.,][]{Sanchez2021, Barrera2021}. To account for the scenario in which the correlations between residuals could have a statistical origin, we derive them using a set of mock data based on the best fit of each scaling relation provided in the previous section. 

%that we will find between the different residuals have a possible physical origin, or on the contrary, the relations have an origin related to the stochastic distribution of the data, in this Section we create the scaling relations described above using random data. In the next Section we will compare the behavior of the residuals of these relations with respect to those observed and we will try to find out how reliable our analyzes are. That is, to investigate if the relations that we report from here on are produced by physical phenomena or are the result of stochastic errors.

%The way in which we are going to explore in the following Sections the possible impact of the interaction or merger of galaxies on the rate of star formation, is through the residuals of the scaling relations that we described in the previous Section. 

%We generate the random extensive and intensive properties separately. 
%That is, we do not get the intensive properties by dividing the extensive properties by a random area. We use the same procedure that we will describe below to generate the extensive ones and independently the intensive ones. We will call this random data \textit{mock-data}, but because of the way we generate this random data, it is not formally a numerical simulation that explains physical properties. 
%Here we intend to generate mock galaxies with values of \Sstar, \Smol, and \Ssfr.
We follow a similar procedure as the one described in \citet{Sanchez2021}. First, we generate a distribution of \Sstar\ for 5000 targets. This distribution is designed to be similar as the one presented by the observed galaxies and is randomly populated. Then, using the best-parameters of the SFMS presented in Sect.~\ref{subsec:characterizing_SFG} we derive an estimation of \Ssfr. In a similar way, we estimated \Smol\ using the best fit of the MGMS. Using these estimations of \Smol\ we also derive the \Ssfr\ from the best fit of the SK law. In order to include the uncertainty given by the systematic error of the observations, we perturb these estimations of \Ssfr\ by their typical uncertainties (i.e., $\sim$0.2 dex). A similar procedure applies to the \Sstar\, and \Smol\ with typical uncertainties of $\sim$0.15 dex and $\sim$0.28 dex, respectively  \citep[see,][]{Barrera2021}. 

As expected, the mock-data are closely distributed around the best fit from the observed data set. However, these have
%We note that the mock-data has  
a slightly smaller spread than the observed data, which may be due to the uncertainties assumed to create the synthetic data. 
%These data present a distribution very similar to that estimated with the observed data, that is, the scaling relations are fulfilled.
Using these mock data, we perform the same procedure as in Sect.~\ref{subsec:characterizing_SFG} to derive the best-fit parameters for each of the scaling relations. Finally, as for the observed data set, we derive the residuals of the mock data using their best-fit relations. In the next section we used the distributions derived between the residuals of this mock data to compare them with those derived from observations. These comparisons provide us with a measurement on how similar (or different) is the distribution of the residual for galaxies in different interaction/merger stages with those expected from a statistical origin.

\section{Results}
\label{sec:Results}

\subsection{The $\Dsk - \Dmgms$ diagram} \label{subsec:dsk_dmgms}
\begin{figure*}[ht!] 
\centering
    \includegraphics[width=0.9\linewidth]{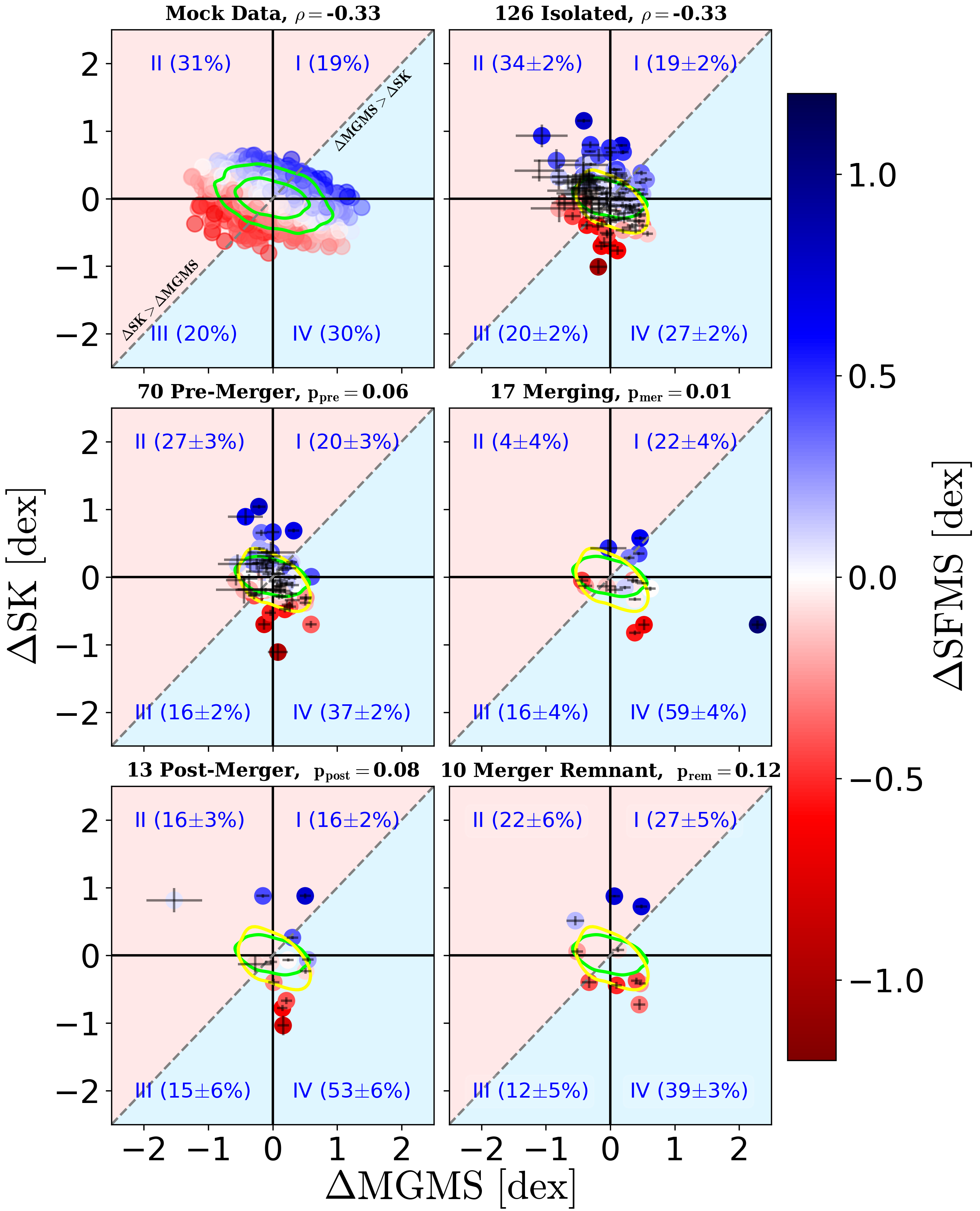}
    \caption{ Distribution of the residuals  of the explored relations: $\Dsk-\Dmgms$ color coded by the residuals \Dsfms. Each panel indicates the number of galaxies studied in the different merging stages, where a circle represents the residual derived in the central region of each galaxy, approximately within 1 Reff (see Sect.~\ref{sec:data_analysis}). The $p$-value of the two-dimensional Kolmogorov-Smirnov test (KS test) is represented by $\mathbf{p}$. And $\mathbf{\rho}$ represents the Spearman correlation coefficient between the datasets shown in the upper panels.
    Green contours in the top-left panel  correspond to the areas encircling $\sim95\%$ and $\sim68\%$ of the mock-data.  Yellow contours in all other panels correspond to the areas encircling $\sim68\%$ of the observed data. Each panel is divided into four quadrants: I, II, III and IV, in parentheses the fraction of galaxies that are located in said quadrant is indicated.  The light red area indicates that galaxies within it have a \Dsk\, greater than their \Dmgms.  The sky blue area indicates that galaxies within it have a \Dmgms\, greater than their  \Dsk.
    }
    \label{fig:stages}
\end{figure*}

%%%%%%%%%%%%%%%%%%%%%%%%%%%%%%%%%%%%%%%%%%%%%%%%%%%%%%%%%%%%%%%%%%%%%%%%%%%%%%%%%%%%%%%%%%%%%%%%%%%%%%%%%%%%%%%%%%%%%%%%%%%%%%%%%%%%%%%%%%%%%

The residuals of the explored scaling relations quantify either the relative enhancement or diminish of an explored property (e.g., the variation of the \Ssfr\ with respect to both \Sstar\ and \Smol, or the variation of \Smol\ with respect to \Sstar). 
%Furthermore, we find that the residuals of the scaling relations using intensive quantities are analogs of the well-known quantities that trace the star-formation activity (sSFR, SFE) and the gas fraction.
Using these residuals, we want to explore the actual impact of the merging process in the star formation activity as well as what could be the main drivers of this activity.

In Fig.~\ref{fig:stages} we plot the residuals of the SK relation (\Dsk) against the residuals of the MGMS (\Dmgms) color coded by the residuals of the SFMS (\Dsfms). In each panel, we plot the different merger stages explored in this study. Top-left panel shows the mock-data sample whereas the top-right panel shows the results from the isolated sample. From left to right and from top to bottom, the other panels show the galaxies in different interaction/merging stages in $\Dsk-\Dmgms$ plane. We divide each panel into 4 quadrants (indicated by: I, II, III, IV). The goal of this division is to compare the fraction of the different samples into each of those quadrants (denoted by the fractions in parenthesis). % We se and in parentheses we indicate the percentage of  galaxies located in said quadrants.  
%The objective of these divisions is to provide a more quantitative characterization of the variations in the properties of galaxies in different merger stages. %, compared to our control sample of non-interacting (isolated galaxies). 
%

The green (yellow) contours enclose $\sim 95\%$ and $\sim 68\%$ of the distribution of the mock (isolated) sample.
%whereas the yellow contour enclose $\sim68\%$ of the distribution of the isolated sample.
When we compare the green and yellow contours encircling  $\sim 68\%$, we find that both samples cover a similar area in the $\Dsk-\Dmgms$ diagram. %
We also find that both samples, mock and isolated, show similar Spearman correlation coefficient for these two parameters ($\mathrm{\rho \sim-0.327}$, and $\mathrm{-0.331}$, respectively). In addition, the Pearson’s chi-squared (applied to the mock data) and Fisher’s exact tests (applied for the isolated sample) for contingency tables resulted in $p$-values of 0 and 0.01, respectively. 
In contrast to previous studies \citep[][]{Ellison2020},  this suggests that the relations between \Dsk\ and \Dmgms\ may not have a physical origin \citep[e.g., ][]{Sanchez2021}. In other words, these relations between residuals could be induced by the stochasticity of the data rather than a physical driver. 
We further quantified the similarity due to between the mock and isolated distributions by deriving the two-dimensional Kolmogorov-Smirnov test (KS test) $p$-value (p=0). Although this value suggests that these two samples are not drawn from a parent sample, we find that the isolated sample has similar fractions for each of the quadrants. This is further evidence that these two samples are similar to each other. We also note that there are significant outliers in the isolated sample in comparison to the mock data.
For instance NGC~3773 (located at -0.4, 1.1) and NGC~5630 (located at -0.9, 0.9) show an increase in their SFE and in their sSFR, but a decrease in their \fg.
Using the two-dimensional $p$-value we compare each interacting/merging sample with the isolated one. We find the highest $p$-value for the  merger remnant galaxies ($\mathrm{p_{rem}
=0.12}$) suggesting that both samples may be extracted from a single two-dimensional distribution. For the other samples, we find that exhibits lower $p$-values, close to zero ($\mathrm{p_{pre}=0.06}$, $\mathrm{p_{mer}=0.01}$ and $\mathrm{p_{post}=0.08}$ for the pre-merger, merging and post-merger samples, respectively). This indicates that such samples are significantly different from the isolated sample.
%
%whereas the sample of active mergers exhibit the lowest $p$-value  ($\Pvi=0.01$) indicating that both samples are significantly different. For the other two merging stages, the results are not conclusive regarding whether their samples and the isolated one are derived from the same parent sample ($\Pvi=0.17$ and 0.15 for the pre-merger and merger remnant, respectively).
%
Regarding the location of the interacting/merging galaxies in each of the quadrants, we find that, unlike the control sample, a significant fraction of galaxies have a diminished in their SFE traced by their \Dsk\, (i.e., most of the galaxies are located in the quadrants III and IV). This is significantly evident for the merger and post-merger samples ($ \mathrm{\sim75\pm6\%}$ and $\mathrm{68\pm8\%}$, respectively). From these galaxies in these two quadrants, we also find that a significant fraction of galaxies (in comparison to the isolated sample) is located in the quadrant IV (i.e., where galaxies having a diminish in SFE but an increase in \fg). It is interesting to note that in our sample we find a few outliers. For example, the interacting galaxy NGC~7253B located in the quadrant IV. This galaxy, although has a negative value of \Dsk, shows both the largest \Dmgms\ and \Dsfms\, from the entire sample. This interacting galaxy and its pair (NGC~7253A), show evident dust lines in their center as well as blue tails and plumbs.  
From this analysis, we find that interacting galaxies, in comparison to the isolated sample, tend to have a diminish of their SFE in their centers, traced by the negative values of \Dsk. However, we find a significant enhancement in their gas fraction, traced by their \Dmgms.
%
%This analysis suggests that the interaction/merging mechanisms affect the amount of central gas in galaxies. 
%%%%%%%%%%%%%%%%5
This analysis suggests that rather than having an impact in the increment of the central star-formation, the interaction/merging mechanisms affect the amount of central molecular gas in galaxies. 
\subsection{Main drivers of the excess/deficit of SFR}
\label{subsec:driver_SFR}

\begin{figure*}[ht!] 
\centering
    \includegraphics[width=0.97\linewidth]{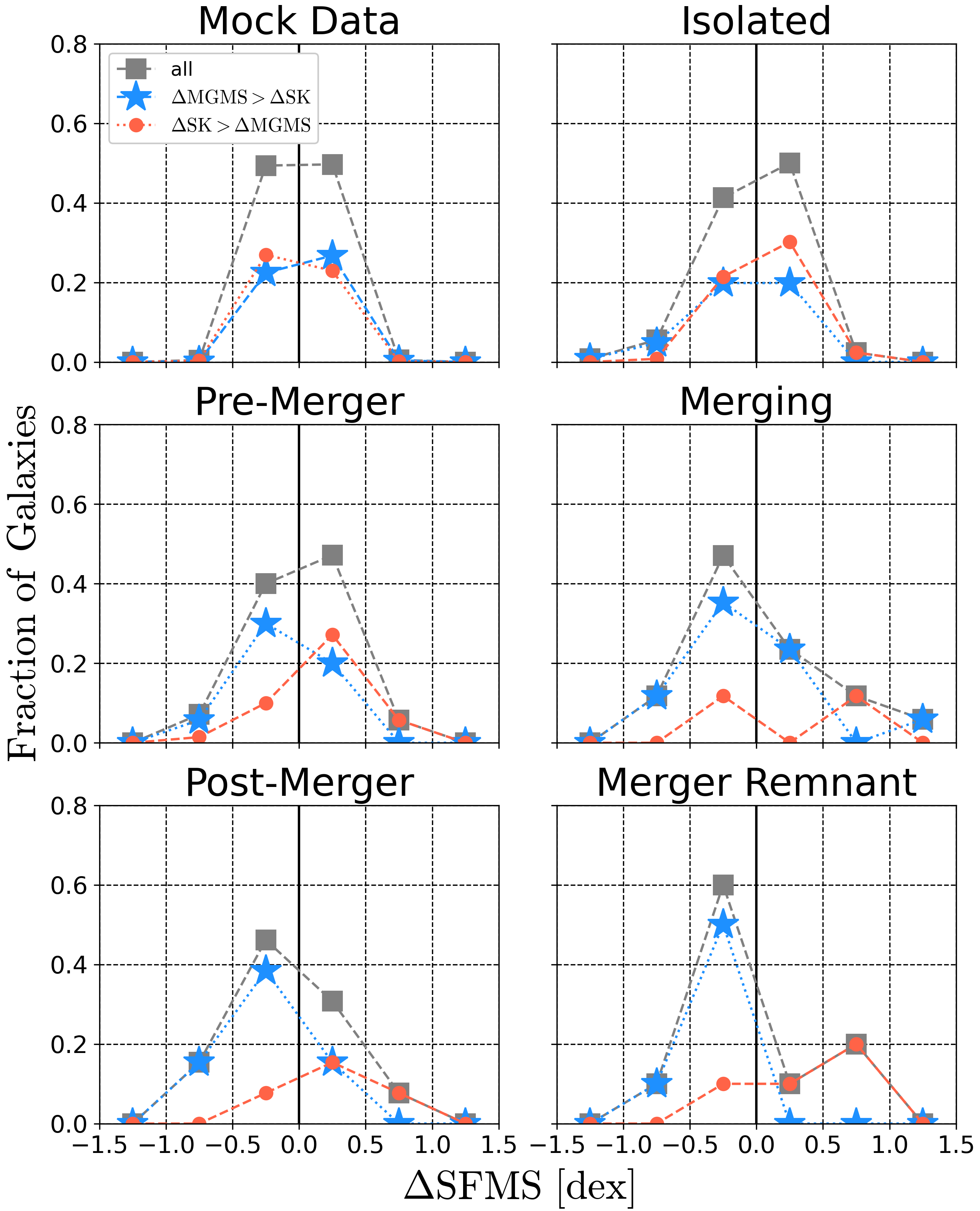}
    \caption{Fraction of galaxies in bins of  \Dsfms\, for each different merger stage. Grey squares show the total number of galaxies per bins. Blue-stars indicate those galaxies in which \Dmgms\, is greater than \Dsk\, (located in the light blue area of Fig.~\ref{fig:stages}), thus, objects with a larger excess in \fg\, than SFE with respect to the average population.  Red-circles indicate just the opposite, i.e., galaxies in which \Dsk\,  is greater than \Dmgms\, (located in the light red area of Fig.~\ref{fig:stages}), thus, objects with a larger excess in  SFE than \fg\, with respect to the average population.
    %Distribution of the residuals of the SFMS relation (\Dsfms) by bins for each merger stage. The stars indicate that in these galaxies, the molecular gas fraction (\Dmgms) is greater than the star formation efficiency (\Dsk), which are located in the light blue area of Fig.~\ref{fig:stages}. Circles indicate the opposite: the SFE is greater than the \fg, these galaxies are located in the light red area of Fig.~\ref{fig:stages}.
}
    \label{fig:stages_Eff_fuel}
\end{figure*}
%So far, we have explored how the merging event affects the SFE and the gas fraction, traced via \Dsk, and \Dmgms, respectively. 
%Here, using as gauge of the star-formation activity the sSFR or equivalently \Dsfms, we study the possible mechanism(s) that could drive this activity in merging galaxies by exploring the interplay among these three parameters. 
%
%To do so, we start by shading in Fig.~\ref{fig:stages} with light red and light blue those regions where  \Dsk\ greater than their \Dmgms, and the opposite, respectively.
%
%Using this information in Fig.~\ref{fig:stages_Eff_fuel}, for each of the sub-samples used in this study we plot for a given range of \Dsfms, the fraction of galaxies with $\Dsk>\Dmgms$, and $\Dmgms>\Dsk$ (red circle and blue star, respectively). 
%
%For the mock-data we find that $\mathrm{\sim50.0\%}$ of the simulated galaxies have an increase in their \Dsfms\ with half of the sample has $\Dsk>\Dmgms$. As expected, simulated galaxies have the same fraction with positive and negative values of \Dsfms\ and that for each of those samples, the fractions of galaxies with $\Dsk>\Dmgms$ and $\Dmgms>\Dsk$  is similar. Thus,  in the central region of these simulated galaxies, it is not possible to distinguish whether star formation enhancements are mainly driven by efficiency (a large fraction of $\Dsk>\Dmgms$) or by the amount of molecular gas (a large fraction of $\Dmgms>\Dsk$). In other words, both parameters are equally important in setting the star-formation activity.

In the previous section, we explore how the merging event affects the SFE, and \fg, traced by the \Dsk\, and \Dmgms, respectively. In this section we explore the dependency of the excess/deficit in SFR for a given stellar mass (gauged by \Dsfms) with a possible excess/deficit on the other two explored parameters for the different merging stages. To achieve this, we compare the fraction of galaxies in different bins of \Dsfms, segregated into two groups: ($i$) those galaxies with an excess of \fg\, with respect to SFE (i.e., $\Dmgms>\Dsk$), and ($ii$) those galaxies with an excess of SFE with respect to \fg\, (i.e., $\Dsk>\Dmgms$). %The first group would correspond to galaxies in which there is a larger increase of the molecular gas with respect to the star-formation efficiency, while the second represents the opposite.  
Following \cite{Moreno2021}, comparing the fraction of galaxies between ($i$) and ($ii$), for a given bin of \Dsfms, we could suggest a possible main driver for that variation of SFR. Thus, if the fraction of galaxies in ($i$) is larger than the fraction in ($ii$), we suggest that the main driver for that enhancement (deficit) of SFR is fuel (efficiency) driven. On the contrary, if the fraction ($i$) is smaller than the fraction ($ii$) we consider than the SFR enhancement (deficit) is efficiency (fuel) driven.       

%By exploring the distribution as a function of \Dsfms\, it is possible to determine what is the strongest driver for the enhancement or decrease of the SFR in different merging stages.
In each panel of Fig.~\ref{fig:stages_Eff_fuel} we show the fraction of galaxies at different bins of \Dsfms\ (gray squares) for each merging stage (including the mock data). In each of these bins, we also show the fraction of galaxies with $\Dmgms>\Dsk$, and  $\Dsk>\Dmgms$  (blue stars and red circles, respectively).
%without making the above separations (grey squares).the fraction of galaxies with $\Dmgms>\Dsk$, and  $\Dsk>\Dmgms$ (blue star and red circle, respectively) for different bins of \Dsfms. 
%
For the mock-data, we find that $\mathrm{\sim50.0\%}$ of the sample has an increase in their \Dsfms. Furthermore, for this fraction of the sample we find that half of it has $\Dsk>\Dmgms$. Given the fact that we build this mock sample by adding random values to the best scaling relations, it is expected to find similar fractions of galaxies for groups ($i$) and ($ii$).
%As expected, simulated galaxies have the same fraction with positive and negative values of \Dsfms\ and that for each of those samples, the fractions of galaxies with $\Dsk>\Dmgms$ and $\Dmgms>\Dsk$  is similar.
Thus, in the central region of these simulated galaxies it is not possible to distinguish whether star formation enhancements are mainly driven by efficiency (a large fraction of $\Dsk>\Dmgms$) or by the amount of molecular gas (a large fraction of $\Dmgms>\Dsk$). In other words, both parameters are equally important in setting the star-formation activity.
In comparison to the mock sample, we observe a similar behavior in the isolated sample (top middle panel of  Fig.~\ref{fig:stages_Eff_fuel}). $\mathrm{\sim52 \pm3\%}$ of the control sample has an increase in their \Dsfms. For these galaxies, more than half of them present a $\Dsk>\Dmgms$ ($\mathrm{\sim62\pm3 \%}$). Although this could indicate that the enhancement in SFR for the isolated galaxies is efficiency-driven (see the scheme above), we acknowledge that the fraction of galaxies with $\Dmgms>\Dsk$ is  similar ($\mathrm{\sim38\pm3\%}$), preventing us to significantly constrain the main driver of SFR enhancement in this sample. Similar conclusions can be drawn from isolated galaxies with negative values of \Dsfms. These results suggest that for the sample of isolated galaxies it is not evident what parameter drives the diminishment or enhancement of star formation activity. These results suggest that these three parameters may not strongly correlate to each other for isolated star-forming galaxies \citep[][]{Sanchez2021, Barrera2021}. 

%On the other hand, for those isolated galaxies with $\Dsfms > 0$, we find a slightly large fraction of galaxies with $\Dsk>\Dmgms$ ($\mathrm{\sim 32.6\% \,vs\, \sim 21.4\%}$) suggesting that the parameter driving the star formation  appears to be the efficiency. However, given the small differences in the fraction of isolated galaxies with positive and negative values of $\Dsfms$, 

%That is, for both samples whether they present an increase or decrease in their sSFR, there is no main driver of this activity. In other words, in the central region of these galaxies, star formation can be driven by efficiency or by the amount of molecular gas.

%Similar to the control samples, 
In the pre-merger sample (top right panel of  Fig.~\ref{fig:stages_Eff_fuel}), we find a similar fraction of galaxies with an enhancement in sSFR ($\mathrm{\sim53\pm4\%}$ with \Dsfms $> 0$) in comparison to the control sample. Furthermore, these galaxies with \Dsfms $> 0$ have a similar fraction of galaxies with $\Dsk>\Dmgms$ ($\mathrm{\sim 62\pm4\%}$) in comparison to the isolated sample. For galaxies with $\Dsfms < 0$ ($\mathrm{\sim47\pm4\%}$), we find that, unlike the isolated sample, the fraction of galaxies with $\Dmgms>\Dsk$ is significantly larger ($\mathrm{\sim76\pm4\%}$).
% ($\mathrm{\sim 37.1\% \,vs\, \sim 11.4\%}$)
%
Thus, for galaxies in pairs, our results suggest that there is no significant increase in the fraction of galaxies with an enhancement in their SFR in comparison to the isolated sample. Furthermore, for companions with an enhancement in their SFR, we are not able to clearly disentangle the possible driver of these enhancements as we find that those galaxies present a similar fraction of galaxies with $\Dsk>\Dmgms$ and $\Dmgms>\Dsk$. This is similar to what we find in isolated galaxies. However, for companions with a diminish of SFR, we find a large fraction of galaxies with $\Dmgms>\Dsk$ suggesting that the lack of star formation is efficiency-driven. This is that, even though there is a large fraction of galaxies with an increment in their gas fraction there is no enhancement in their star formation. 

%This suggests that for pairs of galaxies, on the one hand, a low efficiency could be the responsible for having companion  galaxies with a diminish in the star formation.

%On the other hand, a high star formation efficiency in pairs of galaxies could be responsible for the increment of \Dsfms. In other words, the efficiency, rather than the gas fraction, could be the parameter that modulates the star formation activity in these galaxies. 

For the other merging stages (merging, post-merger and merger remnant) we find that a large fraction of galaxies have a deficit of SFR, in contrast to isolated and pre-merger stages where the samples have a similar fraction of galaxies with an enhanced and deficit of SFR ($\mathrm{\sim59\pm7\%}$, $\mathrm{\sim62\pm7\%}$ and $\mathrm{\sim70\pm7\%}$, respectively). For these galaxies with  $\Dsfms < 0$, the fraction of galaxies with a $\Dmgms>\Dsk$ is greater than for isolated galaxies, and this fraction increases at each stage ($\mathrm{\sim80\pm7\%}$, $\mathrm{\sim88\pm7\%}$, $\mathrm{\sim86\pm7\%}$ for the merging, post-merger and merger remnant, respectively). These results derived from the samples at different interaction/merging stages suggest that the merger event induces a decrement (rather than an increase) in the central star formation activity in comparison to the isolated sample. Furthermore, for those galaxies with a decrement in the star-formation activity we find that the majority of those have an excess in their gas fraction in comparison to their efficiency. According to the scheme presented above, this indicates an efficiency-driven halt of the star formation due to the merging process. 
  
In general, our results suggest that for interacting/merging galaxies, the decrease (increase) in star formation, traced by \Dsfms, is driven by the inability (ability) of the molecular gas to be efficiently transformed into new stars, rather than the amount of raw material available to form those stars. In other words, the efficiency seems to be a significant driver in shaping the star formation activity across the interaction/merger event.
%
%In galaxies with an increase in their sSFR we observe that in the active merger sample the main driven is  molecular gas ($\Dmgms>\Dsk$), but in the post-merger and remnant samples the main driven is efficiency ($\Dsk>\Dmgms$.
%

%same thing, but only in galaxies that have an increase in their sSFR, in galaxies that have a decrease, there is a greater fraction where $\Dmgms>\Dsk$, that is, efficiency is the main driver in suppressing star formation.

%\vspace{15mm}

%%%%%%%%%%%%%%%%%%%%%%%%%%%%%%%%%%%%%%%%%%%%%%%%%%%%%%%%%%%%%%%%%%%%%
%-------------Discussion Y Conclusions
%%%%%%%%%%%%%%%%%%%%%%%%%%%%%%%%%%%%%%%%%%%%%%%%%%%%%%%%%%%%%%%%%%%%%

\section{Discussion}
\label{sec:discusion}

%\subsection{Comparison with previous observations}
%\label{sec:Comparacion_observaciones}

In this study, we explore the impact of galactic interactions/mergers on the specific star formation rate (sSFR), molecular gas fraction (\fg) and star formation efficiency (SFE), as well as the possible driving mechanism of star formation, in the central region of CALIFA galaxies. 
%These galaxies have CO observations provided by CARMA and APEX, which were integrated into APEX beam coverage ($\sim 1~\Reff$).  We divided our interacting/merging sample into four stages to compare with our sample of isolated galaxies. This study explores the largest sample of galaxies with different merger stages having molecular gas data and IFS optical data so far.
%
%As we described in the previous section our results suggest that the greatest impact of galactic interactions and mergers is the accumulation of molecular gas at the center of galaxies rather than an increase in their star formation. We also found that the main driver of star formation appears to be efficiency rather than the amount of molecular gas.
Similar studies with smaller samples and/or heterogeneous data set,  have been previously carried out. \citet{Pan2018} studied a sample of 58 interacting galaxies in pairs ($\mathrm{r_{p}<70~kpc}$) with measurements of molecular gas from four different surveys (and different beam coverage): JCMT PI, JINGLE, JINGLE and xCOLD GASS \citep[see][]{Pan2018}. They used the global SFR and \Mstar\, estimations taken from the MPA-JHU DR7 catalog.
%
%Assuming that CO and SFR distributions trace well between them in nearby galaxies \citep{Leroy2009}, the authors perform exponential molecular gas distribution with a  profile that follows that of stellar light \citep{Saintonge2012}
Assuming a Schmidt-Kennicutt relation between the global SFR and the expected CO emission \citet{Leroy2009,Saintonge2012}, they re-calibrated the observed CO measurements (for a given beam coverage) to a global CO estimation. 
They found that a significant fraction of galaxies in pairs exhibit an increase in their SFR and gas fraction in comparison to a control sample. Although we find that mergers and post-merger galaxies indeed show an increase in their gas fractions in comparison to our sample of isolated galaxies, we do not find significant differences in our sample of pre-merger (pairs) galaxies. We also note that, contrary to this study, we do not find a significant increment in the SFE (traced by \Dsk) between our interacting/merger sample at different stages and the control sample. These differences between our results and those provided by \citet{Pan2018}, could be due to different factors. On the one hand, our CO measurements come from homogeneous observations (i.e., the same beam coverage) provided by both the resolved CARMA and integrated APEX observations. Furthermore, thanks to the IFU CALIFA data set, we are able to provide direct measurements of the \Ha\, flux at the same aperture of the CO observations. Thus, different assumptions in the estimations of the CO values and SFR could lead to increments of the SFE and the gas fraction as presented by \citet{Pan2018}. On the other hand, combining our  pre-merger and merging stages as interacting galaxies in pairs \citep[following the classification by][]{Pan2018}, we have a larger sample of galaxies ($\sim$ 61\% more, 87 vs. 54 galaxies in pairs). This could also lead to differences in the main conclusion of our results in comparison to those reported by \citet{Pan2018}.

Using a sample of 11 interacting galaxies drawn from the xCOLD-GASS \citep{Saintonge2016} survey, \citet{Violino2018} also studied the variations between the SFR and the gas fraction. As a control sample they used a set of non-interacting galaxies also drawn from this survey.
Although they used the optical properties derived from the MPA-JHU catalogue, unlike \citet{Pan2018}, they corrected the SFR to match the aperture of the IRAM millimeter telescope data. \citet{Violino2018} found that for their interacting sample with $\mathrm{r_p<30\, kpc}$, (in our classification pre-merger and merging galaxies), the variations in SFR and gas fraction are similar to those derived from the control sample. They suggest that both interactions and internal processes can lead to variation in molecular gas and SFR in the center of galaxies. Although we find similar results for our sample of pairs of galaxies, we find a significant increment in the gas fraction for our sample of merging galaxies. These differences may be due to the differences in samples and measurements as those described above. On the one hand we use a homogeneous data set and on the other hand, our sample is significantly larger than the one presented by \citet{Violino2018}. 
%
%a sample of 11 xCOLD GASS galaxies with SFR derivations from the MPA/JHU catalogue. Unlike the \citet{Pan2018} study, its correction was made to the SFR at the aperture of the IRAM millimeter telescope. \citet{Violino2018} suggest that for their sample with $\mathrm{r_p<30\, kpc}$, (in our classification pre-merger and active-merger), both interactions and internal processes can lead to a increase in molecular gas and SFR. 
%
%These results are consistent with our results from the pre-merger sample but active-merger sample we found a larger fraction of galaxies with increments in \fg\, and decrement in sSFR. 
%
%One possible explanation for this difference is that as mentioned by \citet{Pan2018}: your results are not applicable for all merging galaxies. In addition to the different models used to calculate the conversion factor \aCO\, in these studies \citep[see][]{Violino2018,Colombo2020}.
%but only those galaxies which, through the interaction, gain a boost in their star formation activity.

Using spatially resolved data, \citet{Kaneko2022} studied the behavior of these properties in 4 pairs of galaxies (ARP 84, VV 219, VV 254, and the Antennae Galaxies),
that corresponds to merging galaxies in our classification scheme. They compare these measurements to 11 non-interacting galaxies. The CO maps for these samples were obtained using the Nobeyama radio telescope observations, whereas the SFRs were derived using data from GALEX observations. They found that for integrated properties, neither the sSFR nor the SFE, exhibit significant increments in comparison to the global measurements of isolated galaxies. Although their sample is smaller compared to ours, the results from this study agree with those presented here. 
%One of its results is that globally the sSFR and SFE do not present significant increases (compared to the control sample). 
%At kpc scales there are increases in the molecular gas fraction and peaks in the SFE in certain regions. 
%This results for global properties agree with our results.
%%%%%%%%%%%%%%%%%%%%%%%%%%%%%%%%%%%%%%%%%%%%%%%%%%%%%%%%%%%%%%%%%%%%%%%%%%%%%%%%%%%%%%%%%%%%%%%%%%%%%%%%%%%%%%%%%%%%%%

%\subsection{Comparison with simulations}
%\label{sec:Comparacion_simulaciones}

Besides observational studies \citep[e.g.][]{Ellison2020b,Thorp2022}, there has been simulations to exploring star formation in interacting galaxies, the aim is to investigate whether the enhancement of star formation is due to the amount of cold gas or to efficiency. For example,
\citet{Moreno2021} conducted high-resolution hydrodynamical simulations focused on the central kiloparsec of Milky-Way like galaxies in pairs (with mass ratios from 2.5:1 to 1:1) with projected separations of $\mathrm{20<r_{p }<120\,kpc}$. 
We should remark that, strictly speaking, it is very difficult to compare quantitatively our results with these simulations. On the one hand, our sample has a wide range of morphologies (furthermore, we are unaware of the morphology of the progenitors of post-merger and merger remnants). On the other hand, in contrast to simulations,  we are not able to have a rigorously mass ratio selection for our sample of pairs and merging sample. Furthermore, the physical area used to measure the properties in those simulations is fixed (1 kpc) whereas the area used in this study varies depending on the APEX beam coverage ($\mathrm{\sim5.2\,kpc}$). 
Therefore, the possible discrepancies between our results could be due to these limitations. In particular, we consider that the different spatial coverages in these studies could induce a major discrepancy.
Despite these, qualitative  comparisons can provide guidelines on what to expect in terms of the numerical simulations on merging galaxies. In these simulations, \citet{Moreno2021} found that the percentage of primary galaxies showing an increase in their amount of molecular gas is $85.4\%$. This is in agreement with our results; in our pre-merger and merging samples $\mathrm{\sim57\pm4\%}$ and $\mathrm{\sim81\pm6\%}$ present an increase in their \fg, respectively. However, we should also note that in our samples the percentage of galaxies with increases in their sSFR ($\mathrm{\sim53\pm4\%}$ and $\mathrm{\sim41\pm7\%}$) is lower compared to the results from these simulations ($67.7\%$). 

Furthermore, \citet{Moreno2021} found that for those galaxies with an increment in their SFR (in comparison to the isolated galaxies) the fraction of these with an increment in the amount of gas is larger in comparison to those exhibit an increase in their star formation efficiency, suggesting that the enhancement in the SFR for interacting galaxies is driven by the amount of gas in the central kpc (so called fuel driven enhanced star-formation). In comparison to the analysis from \citet{Moreno2021}, we find mixed results. On the one hand, those simulations show that for those galaxies with enhanced SFR in the pre-merger stage the fraction of galaxies with an enhancement in the \fg\ is larger in comparison to the fraction with an enhancement of SFE. Contrary to the results from the simulations, our analysis suggests, that the enhancement in the SFR for galaxies in pairs is efficiency driven, rather than fuel driven (see top-right panel in Fig~\ref{fig:stages_Eff_fuel}). We also note that a similar, but reduced effect, is observed in the isolated sample: for those galaxies with an enhancement in sSFR the fraction of galaxies with enhanced SFE is slightly larger than those with a \fg\, enhancement. On the other hand, for the merging galaxies, we find that for some bins of positive \Dsfms\ the enhancement in sSFR is fuel-driven. Although not explored in those simulations, we should note that the observed enhancements in sSFR for the remaining stages of interaction are seem to be driven by efficiency instead of the excess of molecular gas. Regarding  galaxies with a decrement in their SFR, \citet{Moreno2021} found that the fraction of these with an increment in the amount of gas is larger in comparison to those exhibit an increase in their SFE, suggesting that the suppressed in the SFR by interacting galaxies is efficiency driven. In this study, we agree with these simulations for the suppressed star formation.
%we found that for those galaxies with decrease sSFR from pre-merger until the merger remnant present the same driven as these simulations, that is,  the suppressed star formation is efficiency driven. 
These results suggest that the primary effect of interactions and mergers rather than increase the amount of gas to enhance the SFR in the center of galaxies is efficiency that drives either an enhancement or a suppression in star formation, although there is an increase in the amount of molecular gas.
%
%These results suggest that the primary effect of interactions and mergers rather than increase the amount of gas to enhance the SFR in the center of galaxies is to ignite the star formation with the already available molecular gas in these galaxies. 
%
%The above is based on the idea that the enhanced shear in galactic centers may result in the solenoidal mode becoming the typical driving mode in those environments.
%Additionally, the solenoidal mode of turbulence driving has the potential to significantly decrease the SFR when compared to compressive driving \citep{Federrath2012,Federrath2016}
More research is needed to try to understand in detail which processes govern the level of SFE to enhance or suppress star formation. For example, there are studies that conclude that the SFR is primarily controlled by interstellar turbulence and magnetic fields \citep{Federrath2012,Federrath2016}. 

%One possible explanation for the low efficiency of star formation could be the solenoidal mode of turbulence driving, as it has the potential to significantly decrease the SFR \citep{Federrath2012,Federrath2016}}.

Including this study, most of the explored properties of interacting galaxies are derived from integrated measurements (either from the central or entire target). However, as spatially-resolved entities, the impact of the interacting and merger can be different at different locations in the galaxy. Thus, spatially resolved observations are necessary to analyze the gas distribution and compare it in galaxies interacting with isolated galaxies. Unfortunately, the studies available using IFS in this field are scarce. 
Using a sample of 31 galaxies (17 pairs and 14 post-merger galaxies) including in the ALMaQUEST sample\footnote{ALMA MaNGA QUEnching and STar formation (ALMaQUEST) survey includes optical IFS observations from  the MaNGA survey and sub-millimeter interferometric observations from ALMA} \citep{Ellison2020}, \citet{Thorp2022} explored how the interaction and merger affects the drivers of the enhancement of star-formation at both integrated and spatially resolved measurements. Their results suggest that for a given interacting galaxy, even though at global scales a given driver (either fuel or efficiency) dominates the star formation, it may no be the same dominant driver at kpc scales. Furthermore, they found that even within a given galaxy different regions could exhibit a different driver for the enhancement of star-formation, and this is also independent on the interaction stage. Although our results are not directly comparable with spatially-resolved measurements, they suggest that at the central region of interacting/merging, galaxies the efficiency appears to be the most significant driver of both the enhancement and diminish of star-formation. In a future study we explore how our explored parameters (\Dsfms, \Dsk\, and \Dmgms) vary at kpc in a sample of interacting/merging EDGE galaxies. 

Finally, we also note that our findings align well with the observations and simulations of the Central Molecular Zone (CMZ) of the Milky Way \citep{Morris1996}.   The CMZ presents a lower star formation rate than expected based on its amount of molecular gas content \citep[e.g.][]{Immer2012,Longmore2013,Barnes2017}. The reasons for this lower SFR are currently being intensely investigated. For example,  \cite{Henshaw2022}  propose that star formation in the CMZ may be linked to the overall macroscopic evolution of the system, which could be driven by discrete episodes of feedback (and/or accretion), and/or by local interstellar medium conditions that raise the critical density for star formation. Consequently, studying the CMZ will provide insight into the physical mechanisms that result in a reduced SFR, allowing us to extrapolate to the extragalactic context and better understand how interactions and mergers affect the evolution of galaxies in the central region.

\section{Summary and conclusions} 
\label{sec:conclusions}

In this study, we explore the impact of galactic interactions/mergers on the star formation using a sample of 236 star-forming galaxies with homogeneous measurements of the optical and molecular gas properties across their central region (at $\sim$ 1~\Reff, i.e., the APEX beam coverage, $\sim26.3$~arcsec) from the APEX-CALIFA survey. We segregate our sample of interacting galaxies (110 targets) into different merger stages (from pre-mergers to merger-remnant). For the entire sample, we derive the best fits of scaling relations using their intensive properties  (i.e., normalizing by the APEX beam area): the star formation main sequence (SFMS), the Schmidt-Kennicutt relation (SK) and the molecular gas main sequence (MGMS). These fits are shown in Appendix~\ref{Appendix_A}. %These intensive scaling relations exhibit similar behaviors when performed with extensive properties. 
We also generate a mock data set using the best fit of these scaling relations (SFMS, SK, MGMS). %These mock data are to compare their behavior with respect to those observed and to investigate whether the relations that we report from here on are produced by physical phenomena or are the result of stochastic errors. 
Using the best fits of these scaling relations, we obtain the residuals for each of these (\Dsfms, \Dsk\, and \Dmgms). As expected, these residuals are similar to the usual ratios that characterize these three scaling relations: the specific star formation rate (sSFR), the star formation efficiency (SFE), and the gas fraction (\fg). %We then compare the distributions of these residuals with the ratios of each of the scaling relations (sSFR, SFE, and \fg), using both the extensive and intensive properties. Here we found that using the intensive properties the residuals and ratios are proportional with a slope close to 1, and have a correlation coefficient close to one: \Dsfms~vs~sSFR\, (r=0.95), \Dsk~vs~SFE\, (r=0.84) and \Dmgms~vs~\fg\, (r=0.98). This is not the case for extensive properties, where we found lower correlations and slopes. These results suggest that, unlike extensive properties, when normalizing the parameters by the observed area (intensive properties) the effect of distance is mitigated when estimating the best fit of a scaling relation.
%For this reason we use intensive properties, also taking into account that these relations hold at  different scales \citep{Sanchez2021}. So our analysis also shows us that the ratios of the observable data are scale independent to the first order.
%
%Although these ratios appear to have a significant correlation with each other: sSFR~vs~SFE, sSFR~vs~\fg\, and SFE~vs~\fg\,   \citep[in agreement with previous results using smaller samples, e.g., ][]{Lin2020,Ellison2020}, using the mock-data we found similar relations between the ratios. This is a strong indication that such correlations between ratios may not have a physical origin \citep[e.g.,][]{Sanchez2021}. 
%
%For example, similar results have been found using spatially resolved data \citep{Sanchez2021}. The above applies equally to the residual ratios: \Dsfms~vs~\Dsk, \Dsfms~vs~\Dmgms\, and \Dsk~vs~\Dmgms. 
%
Based on these residuals, we explore how the variation (excess or deficit) for the different interacting stage compare with the control sample (see Fig.~\ref{fig:stages}).
%construct the distributions of the samples at the various stages and control sample in the plots: \Dsfms~vs~\Dsk, \Dsfms~vs~\Dmgms\, and \Dsk~vs~\Dmgms\ . % These residuals are summarized in the diagram \Dsk~vs~\Dmgms\, colored by \Dsfms\, and shown in . %This diagram is to carry out a comparative analysis between the samples for different merger stages.
%Depending on the location for a given galaxy in these diagrams, we can know how sSFR, SFE and \fg\, vary (i.e., an excess or deficit).
%
Furthermore, to explore what drives the star formation enhancement or suppression at different interaction/merging stages, we plot in Fig.~\ref{fig:stages_Eff_fuel} the fraction of galaxies with $\Dsk>\Dmgms$, and $\Dmgms>\Dsk$ for different bins of \Dsfms.
Based on the analysis presented in the previous sections, compared to isolated sample, we conclude the following:
%
%In addition, these results (the existence of possible spurious relations between the residues or ratios) show us that it is necessary to carry out a comparative analysis between the samples for different merger stages and for the control sample. %
\begin{itemize}
    \item %We find that our samples from pre-merger to merger remnant the fraction of galaxies showing enhanced sSFR decreases as well as those with enhanced SFE.
    %We find that in our samples, from pre-merger to the merger remnant stage, the fraction of galaxies showing enhanced sSFR as well as those with enhanced SFE is lower compared to the fraction of control galaxies.
    In comparison to the control sample, we do not find a significant fraction of interacting/merging galaxies with an enhancement on their sSFR and SFE. However, we do find that interacting/merging galaxies tend to have a large fraction of galaxies with large central molecular gas fraction (see Fig.~\ref{fig:stages}).  
    \item Although we are not able to fully identify the main  mechanism that modulates the star-formation activity in isolated galaxies (i.e., efficiency vs. fuel driven star formation), we find that, in general, for interacting and merging galaxies both suppression and enhancement of star formation appear to be efficiency driven (see Fig.~\ref{fig:stages_Eff_fuel}).
\end{itemize}

%just over half of the isolated galaxies show an increase in their sSFR, in their \fg\, and in their SFE. In stages 1, 2, 3 and 4 the percentage of galaxies that show an increase in their sSFR decreases ($\sim56.0\%$, $\sim42.9\%$, $\sim41.6\%$ and $\sim27.3\%$, respectively). As your \fg\ increases, the percentage increases in stages 1, 2 and 3 ($\sim62.1\%$, $\sim78.6\%$ and $\sim83.4\%$), and in stage 4 it is almost equal to the percentage of isolates ($\sim54.6\%$). The percentage of galaxies that show an increase in their SFE is less than the isolated galaxies in stages 2, 3 and 4 ($\sim35.7\%$, $\sim33.4\%$ and $\sim45.5\%$), in stage 1 it is very close to the percentage of isolates ($\sim51.4\%$). The dominant mechanism to drive star formation is the SFE in isolated galaxies and in stages 1, 3 and 4, in stage 2 the amount of molecular gas dominates.

Our results indicate that the effect of tidal forces due to the interaction/merger increases the amount of molecular gas in the center of the galaxies, gravitational torques reduced the angular momentum of the molecular gas in the galaxies, moving it inwards. Despite this, in contrast to previous results, we do not find a significant increase in the central star formation for interacting galaxies in comparison to the control sample. We suggest that a possible cause on why the central molecular gas has not been yet transformed into new star could be that it still is in a turbulent stage preventing it to cool down.
Further studies using homogeneous spatially-resolved data (such as the EDGE data set) will allow us to better constrain the role of mergers in shaping the star formation across the optical extension of galaxies in the nearby universe. 

\section*{Acknowledgements}

Y.G.S. thanks  C. Espinosa-Ponce  for his support and comments throughout the development of this work. Y.G.S, and J.B.B. acknowledge support from the grant IA-101522 (DGAPA-PAPIIT, UNAM) and funding from the CONACYT grant CF19-39578.

%%%%%%%%%%%%%%%%%%%%%%%%%%%%%%%%%%%%%%%%%%%%%%%%%%
%%%%%%%%%%%%%%%%%%%%%%%%%%%%%%%%%%%%%%%%%%%%%%%%%%

\section*{Data Availability}

The optical parameters used in this study are taken from the CALIFA survey. These parameters are publicly available. Part of the molecular gas mass has been presented in \cite{Colombo2020}. An updated version of that data set will be presented elsewhere and it will be publicly available. 

%The inclusion of a Data Availability Statement is a requirement for articles published in MNRAS. Data Availability Statements provide a standardised format for readers to understand the availability of data underlying the research results described in the article. The statement may refer to original data generated in the course of the study or to third-party data analysned in the article. The statement should describe and provide means of access, where possible, by linking to the data or providing the required accession numbers for the relevant databases or DOIs.

%%%%%%%%%%%%%%%%%%%%%%%%%%%%%%%%%%%%%%%%%%%%%%%%%%%%

\appendix
\restartappendixnumbering %% solo para AASTEX
\section{Intensive scaling relations}
\label{Appendix_A}

\begin{figure*}%[ht!] 
\centering
    \includegraphics[width=1.01\linewidth]{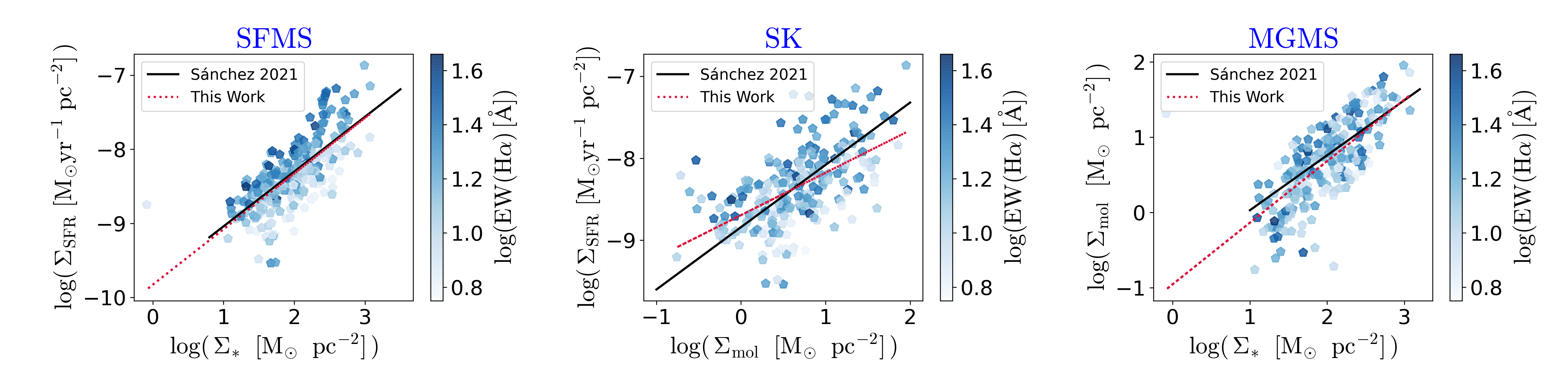}
    \caption{ Distribution of explored galaxies in the \Ssfr–\Sstar\, (left-hand panel), \Ssfr-\Smol\ (middle panel), and \Smol-\Sstar\,(right-hand panel) diagrams. The red lines represent the best fits to the data in our sample (236 star-forming galaxies). The black lines represent the fits of these relations derived by \citet{Sanchez2021}.
    }
    \label{fig:scalling_relations}
\end{figure*}

As mentioned in Sect.~\ref{subsec:characterizing_SFG} of this study, we use the residuals of the scaling relations of the intensive properties (i.e., intensive scaling relations) to quantify the impact of interactions and mergers on central star formation. These intensive properties are obtained by dividing each extensive property (SFR, \Mstar, \Mmol) by the beam area, as described in Sect.~\ref{subsec:derived_parameters}.
Furthermore the scaling relations for these properties are quite similar regardless the observed angular scale \citep[i.e., either integrated or spatially-resolved properties,][]{Sanchez2021}. In each panel of Fig.~\ref{fig:scalling_relations} we show the scaling relations used in this: the star formation main sequence (SFMS), the Schmidt-Kennicutt relation (SK), and the molecular gas main sequence (MGMS). Our best fits (red-dashed lines) are  in good agreement with those presented previously with a similar data set \citep[solid black lines][]{Sanchez2021}. In Table~\ref{tab:fit_scalling_relations} we present the parameters of those best fits.  %are shown: 
%
%We derive the corresponding residuals (\Dsfms, \Dsk\, and \Dmgms, see Sect.~\ref{subsec:characterizing_SFG}) using the best fits of these relations (see Table~\ref{tab:fit_scalling_relations}).
%
%In Fig.~\ref{fig:scalling_relations}  the red lines show our best-fitting linear regression, they are in agreement to those derived by  \citet{Sanchez2021} (black lines).

%

\begin{table}[ht]%[!b]
\centering
%\begin{threeparttable}
  \caption{Best fitting for intensive scaling relations (SFMS, SK, MGMS), $\beta$ intercepts and $\alpha$ slopes. Also, we include the Spearman correlation coefficients  $\rho$.} 
  {\small
  \begin{tabular}{|c|c|c|c|}
   % \hline
    \hline
    \textbf{Relation}  & $\beta$ & $\alpha$ & $\rho$\\
    \hline
    \hline
        SFMS  & $-9.83\,\pm0.10$ & $0.75\,\pm0.05$ & $0.75$ \\
    \hline
        SK & $ -8.70\,\pm0.04$ & $0.52\,\pm0.05$ & $0.59$ \\
    \hline
        MGMS  & $-0.95\,\pm0.11$ & $0.82\,\pm0.05$ & $0.73$   \\
    \hline
\end{tabular}}
\label{tab:fit_scalling_relations}
\end{table}

%%%%%%%%%%%%%%%%%%%%%%%%%%%%%%%%%%%%%%%%%%%%%%%%%%%%%%%%%%%%%%%%%%%%%%%%%%%%%%%%%%%%%%



%%%%%%%%%%%%%%%%%%%%%%%%%%%%%%%%%%%%%%%%%%%%%%%%%%%%

\bibliography{sample631}{}
\bibliographystyle{aasjournal}

\end{document}